\documentclass[a4paper]{jpconf}
\usepackage{graphicx}
\begin{document}
\title{Beamline Instrumentation for Future Parity-Violation Experiments}

\author{Robert Michaels}

\address{Thomas Jefferson National Accelerator Facility, 12000 Jefferson Ave, Newport News, VA 23608 USA}

\ead{rom@jlab.org}

\begin{abstract}
The parity-violating electron scattering community has made
tremendous progress over the last twenty five years 
in their ability to measure tiny asymmetries of 
order 100 parts per billion (ppb) 
with beam-related corrections and systematic errors of a few ppb.   
Future experiments are planned for about an order of magnitude 
smaller asymmetries and 
with higher rates in the detectors.  These new experiments
pose new challenges for the beam instrumentation
and for the strategy for setting up the beam.  In this contribution
to PAVI14 I discuss several of these challenges and demands, with
a focus on developments at Jefferson Lab.
\end{abstract}

\section{Introduction}
\label{Intro}

Parity-violation experiments exploit the fact that a component of the
weak interaction changes sign under a parity transformation, which
isolates the effects due to the weak interaction and provides a tool
to study a variety of physics topics.
In electron scattering experiments, the parity is transformed
by reversing the longitudinal spin, or helicity, of the 
incident electrons.
This method relies crucially on a clean helicity reversal,
such that no other beam parameter, e.g. the angle
or the energy, is affected.  Such effects would cause a systematic
error since the much larger electromagnetic interaction
is very sensitive to these parameters.  

\section{Setting up the Electron Beam}
\label{Beam}

In an ideal electron-scattering parity-violation experiment, 
the two beams corresponding to the two helicity states would
be identical.  In practice, however, imperfections in the laser optics
system at the polarized source will produce some level of coupling of
the helicity to other beam properties.  The enormous efforts to
suppress this coupling at the laser source is described in 
several references (e.g. \cite{humensky,happexprc})
and will not be discussed here.
The experience at Jefferson Lab is that after these efforts
one is left with residual helicity-correlated beam position
differences of typically 50 nm in the beam at the injector which
need further suppression.

\subsection{Adiabatic Damping }
\label{damping}

Linear beam optics in a perfectly tuned accelerator can lead to a
reduction in position differences from the injector to the 
experimental hall due to the 
adiabatic damping of phase space area for a beam
undergoing acceleration~\cite{edwards_syphers}. 
The projected beam size and divergence, and thus the
difference orbit amplitude (defined as the size of the excursion from
the orbit of the design tune), are proportional to the square root of the
emittance multiplied by the beta function at the point of interest.
Ideally, the position differences become reduced by a factor of
$\sqrt{\frac{P}{P_I}}$ where $P$ is the momentum after acceleration and
$P_I$ is the momentum at the injector.  For example, at JLab 
with $P_I = 335$ keV and $P=6$ GeV, the damping factor would be 
$\sim 100$.  This also implies that the region near the injector 
is a sensitive location to measure and apply feedback on 
these position differences, if signals from the beams
of the different experimental halls could be measured separately.

Deviations from this ideal reduction factor can however occur mainly due
to two effects. The presence of XY coupling can potentially lead to
growth in the emittance in both X and Y planes, while a beam
line that does not match well with design, due to imperfections,
often results in growth in the beta function.  Both effects
can translate into growth in the difference orbit amplitude and a 
reduction in the adiabatic damping achieved.
Matching the sections of the
accelerator is an empirical procedure in which the Courant-Snyder
parameters ~\cite{courant_snyder} are measured by making
kicks in the beam orbit, and the quadrupoles are adjusted to 
fine-tune the optics ~\cite{ychao}.  
For future experiments, an investment should be made by accelerator
experts to better understand and automate this procedure 
so that it can be performed efficiently.

\subsection{Spin Flips and Feedback}
\label{flips}

In order to maintain low systematic errors 
there must be at least one, and preferably several,
methods to reverse the helicity.  
The helicity reversals should be uncoupled to other
parameters which affect the cross section.
The rapid, random helicity reversal done 
at the polarized source does a good job of 
reducing the correlation with noise; however, the act of 
changing the sign of a 3 kV voltage on the 
Pockel Cell (PC) to flip the helicity of the laser inevitably
leaves a small residual systematic in the beam.

\begin{figure}[h]
\begin{minipage}{32pc}
\includegraphics[width=32pc]{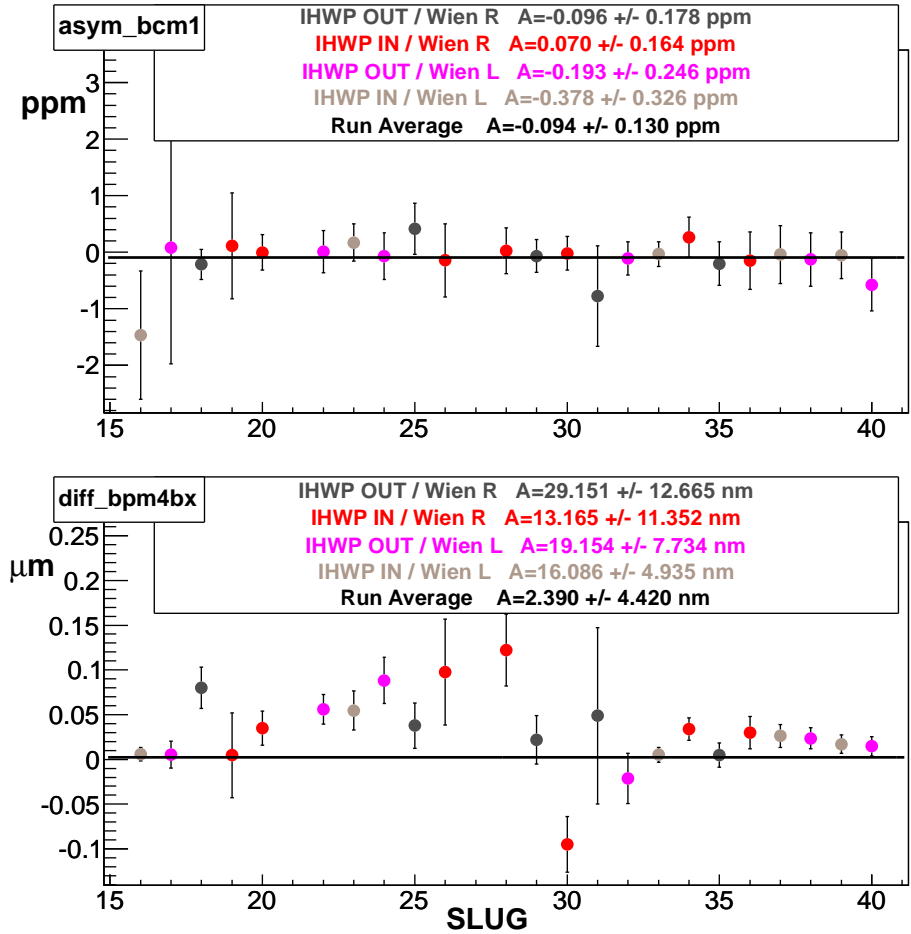}
\caption{\label{label} 
PREX-I \cite{prex} helicity-correlated charge asymmetries (top) and position 
differences (bottom) on a representative monitor versus slug 
(a slug is $\sim 1$ day of running).  The different colors correspond
to four different combinations of insertable halfwave plate (IHWP) 
and Wien used for slow sign reversal.  
To illustrate the systematics, the data points are plotted without sign correction for the helicity flip.  
The final average with 
all sign corrections is shown by the black horizontal bar and was controlled at 
the 5 nm level averaged over the PREX-I run.}
\label{fig:beam_corr}
\end{minipage}
\end{figure}

A standard slow-reversal method 
is a half-wave ($\lambda$/2) plate,
which is periodically inserted into the 
injector laser optical path, reversing the
sign of the electron beam polarization relative to both the 
electronic helicity control signals
and the voltage applied to the polarized source laser electro-optics.
Roughly equal statistics are collected with this waveplate 
inserted and retracted, suppressing 
many possible sources of systematic error.  
In addition to using the ($\lambda$/2) plate, the 
E158 experiment \cite{e158} ran at two beam energies to make 
use of the g-2 precession of spin to achieve another slow reversal.

The $4 \pi$ Spin Flipper is a relatively new apparatus
developed at Jefferson Lab \cite{grames}.  
It is a section of beamline consisting
of two Wien filters with a solenoid magnet sandwiched in between.  
By reversing the spin polarization with the solenoids, rather
than with the Wien filters, a clean flip is achieved  
with minimal change in
the beam trajectory or beam envelope.   
The sign of the B-field in the solenoid determines the beam spin 
direction, while the optical focusing varies, in first order, as
$B^2$ and is therefore insensitive to the sign.

The advantage of having multiple
slow reversals is demonstrated by the 
monitor difference from PREX (fig ~\ref{fig:beam_corr}). 
Note, a ``BPM'' is a beam position monitor, 
and a ``BCM'' is a beam current monitor.
Without the reversals, the 
position differences remained at the $\sim$ 50 nm level (the points without
sign correction) averaged over the experiment; with the reversals,
the differences averaged to the $\sim 5$ nm level (the black lines)
and became a negligible correction 
\cite{prex}.

For the purpose of measuring the sensitivity of helicity correlations
in beam parameters, and for possibly providing feedback to reduce these, 
the injector group has installed a set of fast, pulsed magnets 
at the CEBAF injector.  A dipole magnet can produce helicity 
correlated position or angle differences, while using an arrangement
of weak quadrupole magnets, the spot-size differences at the 
interesting level of $10^{-4}$ can be induced as well
(see section ~\ref{spot1}).
While all parity experiments employ active feedback on the
charge asymmetry, future experiments will likely need active feedback
on other parameters of the beam, 
such as E158 did with piezomirrors in the
polarized source laser optics ~\cite{humensky}.

\section{Normalization}
\label{norm}

When computing the asymmetry from detector signals
one must normalize to beam charge in each helicity state.
In addition, the measured asymmetry must be 
normalized by beam polarization $P_e$
and by a function of the four-momentum transfer $Q^2$.
Here we will neither discuss the measurements of $Q^2$ 
nor the beam polarimetry.
The vital topic of polarimetry was covered at this conference in 
contributions by David Gaskell, Timothy Gay, and Partricia Bartolome.

For the beam charge normalization, the ``noise floor'', or intrinsic
electronic noise of the 
charge monitors, is important.  This is measured by computing the 
difference of charge asymmetries measured by two monitors and computing
the width $\sigma_{\rm diff}$ in the difference.  
At Jefferson Lab, at present, this 
$\sigma_{\rm diff}$ extrapolated
to high currents indicates a noise floor of 60 ppm which
will need to be improved by a factor of 6 for 
the MOLLER experiment \cite{MOLLER}, 
requiring more research on these monitors and the
associated electronics.  
The noise in the BPM monitors and the
intrinsic noise in the beam at 6 GeV both appear to be adequate for
planned experiments, but the beam properties after the 11 GeV
upgrade have yet to be studied.

New digital-signal-processing electronics are being deployed
for the BPM and BCM monitors at Jefferson Lab \cite{musson}
which have the advantages of reduced noise and higher
bit resolution (32 bits).
The electronics process in an FPGA the signals from 
the stripline BPMs or the
cavity BPMs and BCMs.  Consumers of these data will need to
send an electronic gate to define the time of each
integration period and will receive via a fiber optic link 
the digital results, i.e. the integrated charge or position, for 
data acquisition.  The fiber optic link will also have the advantage
of helping to break ground loops.

\section{First Order Beam Corrections}
\label{firstord}

The  scattering rate from a target depends on the five 
``first order'' parameters:  energy, position
and angle of the beam, i.e. $(E, x, y, {\theta}_x, {\theta}_y)$
which are measured by beam monitor
signals $M_i$.
The ``first order'' systematics involve the possible 
helicity-correlations in these parameters and the sensitivities
(derivatives) of the cross section.

\begin{equation}
A = A_D -  A_I - \sum_i {\beta_i \hskip 0.02in \langle \Delta M_i \rangle} 
\end{equation}

Here $A_D$ is the asymmetry in the detector, $A_I$ is the asymmetry
in the beam current, and the $\Delta M_j$ are helicity-correlated
differences in a set of beam monitors $M_i$ which are linearly
related to the beam parameters.  
The coefficients $\beta_j$ are measured during the experiment
by deliberately moving the beam small amounts in all its
parameters, including energy, to measure the sensitivity to these.
Ideally, the corrections to $A_D$ are negligible, and the
systematic errors, which by definition is the error in these
corrections, are also negligible.  That has been the case
for the earlier parity experiments done at JLab \cite{HAPPEXsff, G0sff}
as well as at Mainz \cite{A4sff}.

Future parity-violation experiments \cite{MOLLER,SOLID} at Jefferson Lab
will require BCMs capable of
measuring charge asymmetry widths of 10 ppm at 2 kHz and 
position differences of 3 $\mu$m.
While the Qweak experiment \cite{qweak, qweakNIM}
demonstrated accuracies in beam
monitors close to these values and established an adequately
small jitter in the beam parameters,
more research will be necessary to push the accuracy of
the monitors to robustly satisfy the requirements, 
and the beam properties remain to be
reestablished after the upgrade to 12 GeV 
of the CEBAF accelerator.

\section{Higher Order Beam Corrections and Backgrounds}
\label{HOC}

The next generation of parity-violation experiments will require a 
better understanding and control of the 
``higher order'' corrections, a term used to 
describe both the sensitivity to higher derivatives
of the cross section with respect to beam parameters,
as well as other parameters that are not characterized by
the set of five parameters measured by
the BPM or BCM monitors.  Two examples of these effects are : 1) 
``spot size'' : in this effect, a 
normal-sized beam with no tails has helicity-correlated fluctuations
in its RMS; \hskip 0.05in and 2) ``beam halo" : defined as
significant tails in the beam (many $\sigma$ in radius) 
which can have a helicity component
and which can scrape and cause backgrounds.

\subsection{Spot Size Sensitivity}
\label{spot1}

The effects of spot size can estimated analytically
for a spectrometer experiment like PREX \cite{psouder1}. 
Let $w$ be the RMS width of the beam and $\Delta w$ be the helicity
correlated difference in $w$, with $\Delta w \ll w$.  
For a very narrow (pinhole) acceptance at a distance $D$ from
the target, the accepted angular range is
$\delta \theta = \frac{w}{ D}$.
While the scattering rate $R$ depends strongly on the angle $\theta$
the first-derivative term involving $dR/d\theta$
is suppressed because as the angle increases the 
rate decreases.
The beam-width effect is therefore second-order, and 
 the helicity-correlated differences in the rate $R$ is 

\begin{equation}
\Delta R = \Delta \hskip 0.02in \langle \frac{1}{2} \hskip 0.04in \frac{d^2 R}{d \theta^2}\hskip 0.04in  \frac{w^2}{D^2} \rangle \hskip 0.04in = \hskip 0.04in \frac{d^2 R}{d \theta^2} \hskip 0.02in \frac{ w \Delta w}{D^2}
\label{eq:spotsize}
\end{equation}

For PREX on ${}^{208}$Pb, 
the false asymmetry can be estimated \cite{psouder1} by 
using a typical beam width $w = 100 \mu$m 
and assuming control of 
$\Delta w$ to an accuracy of $10$ nm averaged over a few days,
obtaining a false asymmetry systematic of 2 ppb.
Note that the PREX physics asymmetry is 0.5 ppm and the 
error goal is 3\% which is 15 ppb.  
This establishes a goal to be able to control and measure 
the helicity correlated spot size 
differences $\Delta w$ to an accuracy of about 10 nm averaged 
over a few days.  While this may seem difficult, I note that
comparable accuracies have already been achieved 
for the first order parameters.

\subsection{Measuring Spot Size}
\label{spot2}

While there have been very accurate longitudinal
spot size measurements at the 80 fs level \cite{krafft}, the
measurements of transverse spot size variations, which
use for example wire scanners \cite{freyberg1} or
synchrotron light interferometry \cite{chevstov}, 
have not yet reached a level of accuracy desirable for 
next-generation parity experiments.
Microwave cavity monitors have been proposed \cite{mack1}
as a away to accomplish this.  These monitors are built in a size and shape 
to achieve resonance at a particular eigenfunction 
of the electric field; this provides the desired sensitivity to 
the beam current and the position \cite{slacmonitor}.
Assuming a cylindrical cavity,
the monitor's signal is proportional to $I E \sim J_n(x)$, where $I$ is beam
current, $E$ is the electric field, and $J_n(x)$ is the 
Bessel Function of the first kind,
``n'' denotes the mode, and $x$ is the distance from 
beam center.  A current monitor uses 
the beam to excite the ``n=0'' mode, which is relatively insensitive
to beam position.  A position monitor uses ``n=1'', which to
first order has a linear dependence on the beam position.
A square cavity yields a similar result with sine functions 
as the eigenfunctions.  The question is: Can spot size be determined using 
a higher order mode $n>1$ ?  

If $w$ is the spot size (section ~\ref{spot1})
and the nominal position in the cavity is $x$, the
average of the $n=2$ Bessel Function
is to first order 

\begin{equation}
\langle J_2 (x) \rangle \hskip 0.05in \propto \hskip 0.05in \langle x^2 \hskip 0.04in + \hskip 0.04in x w \rangle
\end{equation}

The signal from such a cavity will be a mixture of 
spot size and position, as well as beam current.
Combining this signal with 
the signals from lower-mode cavities to normalize to 
beam current $I$ and average position $\langle x \rangle$, 
it might be possible to extract the sensitivity to the 
spot size $w$ and its helicity dependence.
Another possible non-invasive method to measure spot size is to use a
stripline position monitor with 8 or 16 poles, which is being 
tried at some facilities \cite{PLi}.

\section{Luminosity and Halo Monitors}
\label{Lumi}

A parity experiment
in an open geometry like Qweak \cite{qweak, qweakNIM} 
or MOLLER \cite{MOLLER} might suffer from dangerous
backgrounds as a result of beam halo.  
The Qweak experiment used a Beam Halo 
monitor to measure and correct for helicity correlations in the halo background.
This detector consisted of a thin aluminum plate surrounded
by lead-shielded lucite detectors and photomultipliers.  

Another tool available for monitoring the beam properties as well
as target boiling is a 
Luminosity Monitor (``Lumi''), which consists of detectors placed symmetrically
about the beam downstream of the target at small angles.
For Qweak \cite{qweakNIM}, four quartz blocks read out 
by lightguides connected
to photomultipliers were used primarily to detector the M{\o}ller
electrons but also served to check the backgrounds from
upstream collimators and beam halo.
The Lumi can also serve to establish a noise floor
since the rates are much higher than in the main
detector and the width of the asymmetry will be narrower.

Future parity experiments will need a
judicious choice of monitoring and collimation to reduce and
to monitor beam halo and other backgrounds.
The experimental apparatus should be simulated carefully and methods to reduce
halo in the accelerated beam, and to make it more predictable,
should be studied.

\section{Acknowledgments}

The author gratefully acknowledges discussions with
Arne Freyberger, Joe Grames, Goeffrey Krafft, Dave Mack, 
John Musson, Mark Pitt, Matt Poelker, Brock Roberts, 
Yves Roblin, and Riad Suleiman.
This work was supported by the
Jefferson Science Associates, LLC,  which operates Jefferson Lab 
for the U.S. DOE under U.S. DOE contract DE-AC05-060R23177.

\smallskip

\end{document}